\documentclass{article}%
\usepackage{amsmath}
\usepackage{amsfonts}
\usepackage{amssymb}
\usepackage{graphicx}%
\providecommand{\U}[1]{\protect\rule{.1in}{.1in}}

\begin{document}

\title{Asynchronous QKD on a Relay Network}
\author{\textit{Stephen M Barnett}$^{1}$\textit{\ }\&\textit{\ Simon JD Phoenix}%
$^{2}\bigskip$\\$^{1}$Dept. of Physics, University of Strathclyde, Glasgow G4 0NG, UK\\$^{2}$Khalifa University, PO Box 127788, Abu Dhabi, UAE}
\maketitle

\begin{abstract}
We show how QKD on a multi-user, multi-path, network can be used to establish
a key between any two end users in an asynchronous fashion using the technique
of bit-transport. By a suitable adaptation of our previous secret-sharing
scheme we show that an attacker has to compromise all of the intermediate
relays on the network in order to obtain the key. Thus, two end users can
establish a secret key provided they trust at least one of the network relays.

\end{abstract}

\section{Introduction}

The elegant and startingly original theoretical idea of Quantum Key
Distribution (QKD) [1] has developed into a mature technology [2] with
commercial systems readily available. Nevertheless, for a variety of reasons,
it still remains something of a curiosity amongst security professionals.
There is a sense in which the technology, however beautiful, addresses a
non-existent problem since the threat models used by security professionals
rarely put key distribution at the top of the list, with good reason. Existing
key distribution mechanisms are considered to be more than adequate to address
the perceived risk. Furthermore, with suitable key-expansion algorithms, there
is little in practice that QKD can achieve that a conventional classical
system cannot.

The security of QKD is based on different principles, however, and
conventional security techniques largely rely on unproven (but reasonable)
assumptions [3]. So, for example, the security proof for a block cipher used
in a suitable mode for key expansion, rests on the assumption that the cipher
is a pseudorandom permutation. Whether we choose a QKD system or a
conventional classical system for our key distribution, we still have to rely
to some extent on our confidence in the underlying principles on which the
security is based.

Another difficulty with QKD is the limitation imposed by the nature of the
technology. The technique relies on the transmission of single quanta, or at
least a reasonable approximation to them. Any network element which is too
lossy, or actively processes the signal in some way, will destroy the
capability of the quantum channel to transmit keys. Thus, installing the
technology on realistic networks poses something of a technical challenge.
Whilst stable and tested solutions exist for point-to-point links, extending
this to a network application is not straightforward and relies on the
introduction of additional trusted network elements to enable the system to
span reasonable distances and to route the signal between the required end
points of the network. Good progress, however, has been made in developing the
basic technique to work on more realistic communication networks [4,5].

The above comments notwithstanding it is likely that QKD will find application
as part of an overall security solution for some situations and networks.
Furthermore, the current threat model to key distribution will significantly
alter as more progress is made towards the development of a working quantum
computer that can process strings of qubits of sufficient size to pose a
threat to existing public-key mechanisms [6,7]. Whilst classical key
distribution techniques based on symmetric cryptography can address the threat
posed by quantum computation, it is by no means certain that these will be an
obvious natural choice over a QKD solution should the need arise for a
widespread overhaul of the existing key distribution techniques based on
public-key cryptography.

In this paper we look at how the bit-transport technique for QKD [8] can be
used on a network in an \textit{asynchronous} fashion to establish keys
between any end-users of the network. The technique requires that the network
relays act as intermediaries to correlate various QKD transmissions together.
We show that with a suitable arrangement of relays an attacker has to
compromise all of the relays on any particular channel in order to obtain the
key. Thus, instead of having to trust \textit{all} of the relays on a channel,
the end users only have to trust \textit{at least one}. We achieve this by a
suitable adaptation of our `drop-out' technique [9] for single QKD channels.

\section{A Single-Relay QKD Channel}

A relay on a QKD channel is used, primarily, to increase the distance. The
conventional way of achieving this is for Alice and Bob to each establish
separate quantum\footnote{We use the term `quantum key' here merely to
describe the process by which the key has been established, that is, by
quantum key distribution. There is, of course, nothing quantum about the key
itself!} keys with the intermediate relay. In an obvious notation Alice ($A$)
establishes a key with the relay ($R$) which we label $QK_{AR}$. Bob
establishes a different key $QK_{RB}$ where we have used the order of the
indices here to denote the `direction' of key establishment which we take to
be the direction in which the quanta are transmitted. The final key, $K$,
between Alice and Bob can be established in a linkwise fashion.

If one relay is not enough to span the distance between $A$ and $B$ with a
quantum key transmission then, clearly, we can use any number of intermediate
relays which each establish a separate quantum key $QK_{R_{j}R_{k}}$. Once all
of the keys between the various network entities have been established, the
final key between Alice and Bob can again be established in a linkwise fashion.

In our previous work [8] we showed how we can use intercept/resend relays [10]
to establish an end-to-end key over an extended distance with no loss of
effective final key size. The primary concern with any relay system is that
the relays have to be trusted intermediaries. In conventional, or
intercept/resend, operation the compromise of a single relay compromises the
entire channel. In [9] we showed how it was possible to modify the
transmission protocol by adding a single relay (at least) so that an attacker
needs to compromise \textit{all} of the relays on a channel in order to obtain
the key. This technique employs the notion of quantum secret sharing developed
for multi-path networks in [11] in which we create distinct logical paths on a
single channel by randomly dropping out the relays from the channel.

\subsection{Asynchronous Bit-Transport on the Channel}

Let us consider a channel over which Alice and Bob desire to establish a
quantum key. Further, let us suppose that a single relay is required to span
the distance so that the channel is of the form Alice $\longrightarrow$\ Relay
$\longrightarrow$\ Bob. There are two ways in which the relay can be operated;
in link-by-link mode, or in intercept/resend mode to establish an end-to-end
key. We have shown how bit transport can be used to establish an end-to-end
key with intercept/resend relays [8], we'll now consider how bit transport can
be used to establish a key between Alice and Bob in an asynchronous fashion
with the help of the relay. The relay establishes an independent QKD channel
with Alice and Bob, respectively. On each channel the bits are sifted with
public announcement of the coding basis and the `bad' channels discarded. At
the end of this process Alice and the relay, and Bob and the relay, possess a
set of data that should, in an ideal world and in the absence of an
eavesdropper, be in perfect respective agreement. The 4 sets of data can be
checked for errors. If the error rate is not too high then the data sets can
be saved and labelled.

After many such transmissions Alice and the relay have $n$ sets of data
$S_{AR}^{(k)}$ and $\bar{S}_{AR}^{(k)}~~(1\leq k\leq n)$, where the bar
denotes the relay's data set which could differ slightly from the
corresponding set of Alice if there are errors on the channel. The relay and
Bob have similar sets of data, $S_{RB}^{(k)}$ and $\bar{S}_{RB}^{(k)}$ and we
assume that they have $m$ such sets. The elements of each set consist of a
tuple $(t,b)$ where $t$ is the timeslot and $b$ is the bit value. So for a
given pair of sets, $S_{AR}^{(k)}$ and $\bar{S}_{AR}^{(k)}$, we would have
elements $(t,b)$ for Alice and elements $(t^{\prime},b^{\prime})$ for the
relay. We would have $t=t^{\prime}$ but in the presence of errors we would
have $b=b^{\prime}$ for most, but not all, of the timeslots. Let us suppose
that Alice and the relay now conduct a secure error-correcting process so that
at the end of this, and with a suitable re-labelling of the timeslot values,
we have that $t=t^{\prime}$ and $b=b^{\prime}$ for all elements. Let us label
the sets after error-correction by $\Sigma$. We further suppose that Bob and
the relay perform the same process on their sets so that they also have
identical sets with elements $\left(  \tau,\beta\right)  $.

Now if Alice and Bob wish to establish a key then the relay can choose one of
the error-corrected sets $\Sigma_{RB}^{(k^{\prime})}$ and one of the
error-corrected sets $\Sigma_{AR}^{(k)}$ essentially at random (or select from
those which initially had a lower error rate, for example). The relay can then
choose the timeslots, at random, from the sets such that $b=\beta$ and simply
announce the respective timeslots to both Alice and Bob. Alice and Bob will
then share the same set of bits which can subsequently be used as a key. An
eavesdropper has to have collected all of the data between Alice, Bob and the
relay in order to have any chance of getting any information about the key,
because she cannot know in advance which sets of data the relay will choose.
Of course, once the timeslots have been announced then Alice and Bob need to
perform a privacy amplification [2,7] on their data in order to eliminate the
possible information the eavesdropper could have gleaned to a negligible
level\footnote{Although privacy amplification can be performed after the
bit-transport (as we have discussed here) this is not optimal because the
linkage of the timeslots gives the attacker a greater number of bits of the
key that she knows with certainty.}.

There are variations on this protocol. For example, the error-correction need
not be done before the linking of the timeslots by the relay (although this
will increase the effective error rate on the final data). The participants on
the channel could, with a collection of error-corrected sets, decide to XOR
these sets together to reduce the potential information of the eavesdropper.
With the necessity to establish only a short key and with a potentially large
number of sets to choose from this technique could reduce the eavesdropper's
information to negligible levels in a similar fashion to the standard privacy
amplification procedure. Alternatively, the privacy amplification could be
done on the error-corrected sets $\Sigma$, individually, before the
bit-transport by linkage of the timeslots. Furthermore, the linkage on this
channel need not be initiated by the relay. Alice, for example, could begin
the linkage by announcing the timeslots she wishes to use which then get
correlated by the relay to suitable timeslots of Bob's. The linkage need not
be restricted to the selection of a single respective set of the participants.
Indeed, elements from different sets could be chosen at random provided that
they have the same bit value. The main limitation of this asynchronous
key-establishment, whichever protocol variation is adopted, is that the relay
has to be entirely trusted by Alice and Bob.

\section{QKD Channels with Multiple Relays}

It is clear that this process can be extended to channels which require
multiple relays. Let us consider a channel between Alice and Bob that requires
2 relays to span the distance. Thus we have a channel of the form :%

\[
A\longrightarrow R_{1}\longrightarrow R_{2}\longrightarrow B
\]

Independent QKD transmissions are run on the channels $A\longrightarrow
R_{1},R_{1}\longrightarrow R_{2}$ and $R_{2}\longrightarrow B$ so that at the
end of many such runs, and after sifting and error-correction, Alice and
$R_{1}$ share $n$ sets $\Sigma_{AR_{1}}^{(i)}$, the two relays share $m$ sets
$\Sigma_{R_{1}R_{2}}^{(j)}$, and $R_{2}$ and Bob share $l$ sets $\Sigma
_{R_{2}B}^{(k)}$. For convenience we shall assume that all of these sets are
of the same size with cardinality $N$. The timeslot index is just an integer
identifying a transmission instance, thus each of these sets consists of
elements of the form $(t,b)$ with $1\leq t\leq N$ and $b\in\{0,1\}$. Each set
is therefore an ordered list of $N$ bit values. An example for $N=10$ is given
below, in which Alice and the first relay have selected\footnote{We assume
that such a selection has been made for the purposes of explanation of the
asynchronous bit-transport technique.} the 4$^{th}$ set from their list of $n$
sets, the relays have selected the 2$^{nd}$ set from their list of $m$ sets,
and the second relay and Bob have selected the 7$^{th}$ set from their list of
$l$ sets. In practice $N$ will be orders of magnitude greater than 10.\bigskip%

\begin{tabular}
[c]{|c|c|c|c|}\hline
$N$ & $\Sigma_{AR_{1}}^{(4)}$ & $\Sigma_{R_{1}R_{2}}^{(2)}$ & $\Sigma_{R_{2}%
B}^{(7)}$\\\hline
1 & 0 & 1 & 0\\\hline
2 & 1 & 1 & 0\\\hline
3 & 1 & 1 & 1\\\hline
4 & 0 & 0 & 1\\\hline
5 & 1 & 1 & 0\\\hline
6 & 1 & 1 & 1\\\hline
7 & 0 & 1 & 0\\\hline
8 & 0 & 0 & 1\\\hline
9 & 1 & 0 & 1\\\hline
10 & 0 & 1 & 1\\\hline
\end{tabular}
\bigskip

Let us suppose that Alice and Bob wish to establish a key of length 4 bits.
We'll consider the case where the relays select the key to be used. $R_{1}$
and $R_{2}$ communicate and agree on 4 elements chosen at random from their
set $\Sigma_{R_{1}R_{2}}^{(2)}$. For example, we suppose that they agree on
the following list (6,1,9,3) giving the key 1101. Now $R_{1}$ chooses, at
random, the indices from the set $\Sigma_{AR_{1}}^{(4)}$ that will give this
key and communicates the list of chosen indices to Alice. So, for example, in
order to communicate the key 1101 to Alice, $R_{1}$ might transmit the list
(3,2,8,9). It is important that once an index value has been selected it is
eliminated from any subsequent choice. The relay $R_{2}$ performs the same
process with Bob and, for example, might transmit the index list (4,9,2,10).
At the end of this process both Alice and Bob will share the key 1101.

As we have noted above, the selection of the bit values need not be restricted
to a single set. These values could be chosen randomly from all available
sets. In this case each bit value index must be accompanied by another integer
which indexes the set from which it is taken. Thus a list of tuples must be
transmitted. So, for example, the first relay could send Alice the list
$[(12,4),(2,1),...(34,10)]$ which would indicate that the first bit of the key
is the 4$^{th}$ element of their set $\Sigma_{AR_{1}}^{(12)}$ and so on.

Of course, more sophisticated schemes for key establishment can be envisaged,
rather than just the straight linkage of the timeslots. For example, Alice and
the first relay could partition the data in their error-corrected sets into 10
bit blocks (say), where the elements of each block are selected at random by
publicly agreeing on a random sequence. In effect, a common random permutation
is applied to the set. The bit values of the key can then be established by
announcing a block and determining the parity. This procedure combines an
element of privacy amplification into the key establishment.

Again, the main limitation on this technique, from a security perspective, is
that each network element knows the final key and so each network element
(that is, the relays) needs to be trusted. If any one relay is compromised
then the key between Alice and Bob can be determined by the attacker. We can
adapt our previous secret sharing technique [9] to alleviate this problem so
that an attacker has to compromise \textit{all} of the relays on the channel.
Let's look at an example of how this works.

\subsection{Securing a Multiple-Relay Channel}

Let us consider the following channel%

\[
A\longrightarrow R_{1}\longrightarrow R_{2}\longrightarrow R_{3}%
\longrightarrow B
\]

Now let us suppose that a successful QKD transmission can be performed
between\ network elements that are, \textit{at most}, 2 steps away. The
following quantum keys between Alice and Bob can therefore be established by
utilizing the bit-transport technique outlined above;\bigskip%

\begin{tabular}
[c]{cc}%
\underline{Quantum Key} & \underline{QKD Channel}\\
$QK_{AR_{1}R_{2}R_{3}B}$ & $AR_{1}R_{2}R_{3}B$\\
$QK_{AR_{2}R_{3}B}$ & $A-R_{2}R_{3}B$\\
$QK_{AR_{1}R_{2}B}$ & $AR_{1}R_{2}-B$\\
$QK_{AR_{1}R_{3}B}$ & $AR_{1}-R_{3}B$\\
$QK_{AR_{2}B}$ & $A-R_{2}-B$%
\end{tabular}
\bigskip

We can see that there is at least one channel in which a given relay does not
participate. So Alice and Bob establish these separate quantum keys and simply
XOR them together to obtain their final key. None of the relays now possess
this final key, and in order to obtain the key an attacker must compromise all
of the relays in the channel. If only one is not compromised, and therefore
trusted, the attacker cannot obtain the final key. For this particular example
Alice and Bob could establish either of the following final quantum
keys\footnote{Of course there are other final quantum keys that can be
established. It is important, however, that any relay has not participated in
the establishment of all of the keys used in the XOR. Thus, for example, if we
tried to establish a key $QK_{AR_{2}R_{3}B}\oplus QK_{AR_{1}R_{2}B}$ then
relay 2 knows the final key and the attacker would only have to compromise
this relay in order to obtain the final key.}:%

\begin{align*}
QK_{AB}  &  =QK_{AR_{1}R_{2}R_{3}B}\oplus QK_{AR_{2}R_{3}B}\oplus
QK_{AR_{1}R_{2}B}\oplus QK_{AR_{1}R_{3}B}\oplus QK_{AR_{2}B}\\
& \\
QK_{AB}^{\prime}  &  =QK_{AR_{2}R_{3}B}\oplus QK_{AR_{1}R_{2}B}\oplus
QK_{AR_{1}R_{3}B}\oplus QK_{AR_{2}B}%
\end{align*}

The second key here would seem intuitively more preferable since relays 1 and
3 only participate in 2 out of the 4 channels. It is obvious how to extend
this to any general network configuration. Indeed, on multi-path networks we
could employ a combination of this single-channel secret sharing and the
multi-path secret sharing developed in [11]. Furthermore, on multi-path
networks we could choose any path, at random, for each key bit we wish to
establish using the bit-transport technique. The attacker would then be in the
position of having to compromise \textit{all} of the relays on the multi-path
network, or collect \textit{all} the data exchanged on all possible paths.

\section{Conclusions}

We have shown how, with a suitable adaptation of our previous bit-transport
and secret-sharing techniques [8,9], an asynchronous quantum key can be
established between any two users on a network in such a way that an attacker
has to compromise all of the intermediate network elements to obtain the final
key. Indeed, in order to obtain even a limited amount of information about the
key the attacker must collect the data between all network elements, even on
different paths. Of course, the standard operating assumptions of a normal
single link QKD channel must be observed. So, for example, the public
communications between the various elements must be authenticated and any
side-channel information must be protected.\bigskip

\textbf{Acknowledgement\bigskip}

SMB thanks the Royal Society and the Wolfson Foundation for financial
support\bigskip

\textbf{References}

\begin{enumerate}
\item Bennett, C.H.; Brassard, G. Quantum Cryptography: Public Key
Distribution and Coin Tossing. \textit{Proceedings of the IEEE International
Conference on Computers, Systems, and Signal Processing}, Bangalore, India,
p.175, 1984

\item Gisin, N.; Ribordy, G.; Tittel, W.; Zbinden, H. Quantum Cryptography,
\textit{Rev. Mod. Phys}., \textbf{2002}, \textit{74}, 145--195

\item See, for example, Katz, J.; Lindell, Y. \textit{Introduction to Modern
Cryptography}; Chapman \& Hall: Florida, 2008

\item Zbinden, H.; Gisin, N.; Huttner, B.; Muller, A.; Tittel, W. Practical
aspects of quantum key distribution. \textit{J. Cryptology,} \textbf{1998},
\textit{11}\textbf{,} 1-14

\item Sasaki, M.; \textit{et.al.} Field test of quantum key distribution in
the Tokyo QKD Network. \textit{Optics Express}, \textbf{2011, }\textit{19}
(11), 10387-10409

\item Shor, P.W. Polynomial-Time Algorithms for Prime Factorization and
Discrete Logarithms on a Quantum Computer. \textit{SIAM J. Comput}.,
\textbf{1997}, \textit{26} (5), 1484--1509

\item Barnett, S.M. \textit{Quantum Information}; Oxford University Press:
Oxford, 2009

\item Barnett, S.M.; Phoenix, S.J.D. Extending the Reach of QKD Using Relays.
\textit{Proceedings of the IEEE GCC Conference}, Dubai, UAE, pp.140-142, 2011

\item Barnett, S.M.; Phoenix, S.J.D. Securing a Quantum Key Distribution Relay
Network Using Secret Sharing, \textit{Proceedings of the IEEE GCC Conference},
Dubai, UAE, pp.143-145, 2011

\item Bechmann-Pasquinucci, H.; Pasquinucci, A. Quantum Key Distribution with
Trusted Quantum Relay.\ \textbf{2005, }http://arxiv.org/abs/quant-ph/0505089
(accessed Mar 19, 2012)

\item Beals, T.R.; Sanders, B.C. Distributed relay protocol for probabilistic
information theoretic security in a randomly-compromised network.
\textit{Third International Conference on Information Theoretic Security}
(ICITS), pp.29--39, 2008
\end{enumerate}

\end{document}